\documentclass[12pt]{article}
\usepackage{epsfig,cite}
\title{Is there a prescribed parameter's space for the adiabatic geometric 
phase?\footnote{Published in Europhysics Letters {\bf 21} (1993) 148-152}}

\bigskip

\author{David J. Fern\'andez C.\footnote{On leave of absence, Departamento 
de F\'\i sica, Cinvestav-IPN, A.P. 14-740, 07000 M\'exico D.F., Mexico} 
\\ Departamento de F\'{\i}sica Te\'orica \\ Universidad de
Valladolid \\ 47011 Valladolid, Spain \\ and \\
Nora Bret\'on${}^\dagger$ \\ Departamento de F\'{\i}sica \\ 
Universidad del Pa\'{\i}s Vasco \\ Apartado 644, 48080 Bilbao, Spain}

\bigskip

\date{October 1992}

\begin{document}

\maketitle

\thispagestyle{empty}

\noindent {\bf Abstract.} The Aharonov--Anandan and Berry phases
are determined for the cyclic motions of a non--relativistic charged
spinless particle evolving in the superposition of the fields
produced by a Penning trap and a rotating magnetic field. Discussion
about the selection of the parameter's space and the relationship
between the Berry phase and the symmetry of the binding potential is
given.

\vskip1.5cm
\bigskip


\vfill\eject

In 1984 Berry [1] discovered a geometric phase, which now is
known as Berry's or adiabatic phase and usually arises when an
eigenstate of the Hamiltonian $H({\bf X}(t))$ evolves cyclically and
adiabatically due to a cyclic and slow change of the parameters
${\bf X}(t)$. That phase was identified as a geometric property of the
space of all possible values of ${\bf X}(t)$ (the parameter space)
[2] and gave rise to the present interest on the geometric
aspects of quantum theory [3--7]. An important generalization
which does not need the adiabatic assumption (or equivalently that the
initial state be an eigenstate of $H({\bf X}(t))$) was proposed by
Aharonov and Anandan [3]. The essential requirement in the
Aharanov--Anandan (AA) treatment is that the system performs a
cyclic motion $|\psi(T)\rangle=e^{i\phi}|\psi(0)\rangle$, then the 
geometric phase is given by:
$$
\beta=\phi+i\int_0^T\langle\psi(t)|{d\over{dt}}|\psi(t)\rangle dt.
\eqno(1)
$$
$e^{i\beta}$ is the holonomy in the projective space of physical
states, not a geometric property of the parameter space. However, as
the adiabatic limit of the AA phase provides Berry phase (BP) this
quantity can be considered also as a geometric property of the
parameter space in this limit.

Recently, attention has been paid to the dynamics of systems
interacting with magnetic fields and the determination of their
cyclic motions [8--9]. For the situations involving
a precessing magnetic field the AA phase and the BP have been
calculated; in all these cases the BP turned out to be a linear function
of $\cos\theta$, where $\theta$ is the semiangle of the cone swept by
${\bf B}(t)$ and is related to the solid angle by
$\Delta\Omega=2\pi(1-\cos\theta)$. To the light of those results one
might conclude that if a precessing magnetic field is involved in
the cyclic evolution of a quantum system then the adiabatic phase
will be simply a linear function of $\cos\theta$. This naive
generalization is wrong because some facts  must be taken into
account. One of them is the ambiguity in choosing the parameter
space when  BP is evaluated according with the original Berry's
definition [1]. Because of the adiabatic nature of BP one could
think that in its determination we only need to consider as forming
the parameter space those quantities depending explicitly on time.
However, in [1] the time--independent parameters were taken
into account to achieve a geometric interpretation of BP (as the
solid angle in the parameter space), but this is not a rule.  For
instance, if we consider a massive charged spherical oscillator with
frequency $\omega_0$ interacting with ${\bf B}(t)$, the BP turns out
proportional to $\cos\theta$ [9], then we don't need to
incorporate $\omega_0$ to the ${\bf B}(t)$--space. Other fact influencing on
the BP is the symmetry of the binding potential. If we choose a
binding potential with less symmetry than a spherical oscillator,
the final BP will change, and this would lead us to conclude that
the old parameter space (the ${\bf B}(t)$--space) must be modified in
order to give a geometric interpretation of those results. These are
the motivations for this paper, where we analize the geometric phases
for a charged spinless particle evolving in the fields produced by
a Penning trap with symmetry's axis along the $x_3$--direction and
simultaneously an orthogonally rotating magnetic field. The system is
described by the Hamiltonian:
$$ 
H(t)= {1\over{2m}}\left({\bf p}+{e\over{2c}}{\bf r}\times{\bf B}(t)\right)^2+{
m\over{2}}\omega_0^2\left(x_3^2 -{{x_1^2+x_2^2}\over 2}\right), \eqno(2)
$$
where ${\bf B}(t)=(B\cos\omega t,B\sin\omega t,B_0), \ e<0$. In order to
determine the cyclic motions, we first classify the system's
parameters according to the confined or unconfined nature induced by
the Hamiltonian (2) on the quantum canonical trajectories. To
simplify this classification, we take particular combinations
of the parameters which for $B=0$ bring the evolution operator
cyclically into the identity operator (the Penning loops
[10], see also [11]). By restricting the parameters of the
Penning loop and the rotating field to the confinement domain, we
find the cyclic states of the system and evaluate their AA
phases and BP. We discuss briefly the physical significance of these phases 
when the system is under the effect of an external periodic perturbation.
 Some remarks about the selection of the parameter
space will be made. Finally we will compare graphically both the
contributions to the BP of the three independent modes in which the
system can be decomposed and $\cos\theta$. So, besides determining
the geometric phases for a physically interesting system we give an
evidence that the simple relationship between the BP and
$\cos\theta$ for the systems previously studied is rather a
consequence of the spherical symmetry possessed by the binding
potential, which is lacking in the problem treated here.

The dynamics of a non--relativistic quantum system is governed by
the Schr\"odinger's equation, which in operator's version reads
($\hbar=1$): 
$$
{d\over{dt}}U(t)=-i H(t)U(t);  \ \ \ \ \ U(0)=I. \eqno(3)
$$
If the Hamiltonian has the form (2), then the evolution operator
$U(t)$ can be obtained by means of the ``transition to the rotating
frame'' (see [12--14]), which maps (3) into a
similar equation but with a time--independent Hamiltonian
$G=H(0)-\omega L_3$. Hence, the solution of (3) has the disentangled
form:  
$$
U(t)=e^{-i\omega tL_3}e^{-iG t}.   \eqno(4)
$$
Note that $H(t)$ and $G$  are quadratic forms of the canonical
6--vector  ${\bf u}^{\rm T}=({\bf r},{\bf p})$. Hence, the general motion's
properties of the system are determined by the kind of symplectic
transformation induced by $e^{-iG t}$ on ${\bf u}$, 
${\bf u}(t)=e^{iG t}{\bf u}e^{-iG t}=e^{\Lambda t}{\bf u}$, where $\Lambda$ is a
$6\times 6$ matrix obtained of $[iG,{\bf u}]=\Lambda{\bf u}$ [12].
For the case in which $\Lambda$ is diagonalizable, we find two
different behaviors, determined by the values taken by the three
dimensionless parameters $\{\alpha={{|e|B}\over{2mc\omega}}, \
\alpha_0={{|e|B_0}\over{2mc\omega}}, \ w={{\omega_0}\over{\omega}}\}$.
In one case all the six roots of the characteristic polynomial of
$\Lambda$, $D(\lambda)=Det(\lambda I-\Lambda)$, are imaginary,
and the spectral decomposition of the symplectic matrix $e^{\Lambda
t}$  will induce confined motions on the canonical trajectories
${\bf u}(t)$. In the other case, at least two roots of
$D(\lambda)$ have a non--null real part, and $e^{\Lambda t}$ will
produce unconfined motions for ${\bf u}(t)$. The above two cases
induce a similar classification on the $\alpha-\alpha_0-w$ space
according to the roots of  $D(\lambda)$ and, consequently, to the
kind of generic motion possessed by the system at each point
$(\alpha, \ \alpha_0, \ w)$. For the sake of simplicity, we will
restrict ourselves to physical situations in which, when $B=0$ the
evolution operator (4) returns periodically into the identity
operator (the Penning loops [10--11]). Moreover, we will take
the simplest Penning loop which appears if we put
$\omega_0={{2|e|B_0}\over{3mc}}$ ($w={4\over 3} \alpha_0$). With this
choice, we have plotted the confinement and unconfinement regions in
the $\alpha-\alpha_0$ diagram of Fig.1. The confinement domain
consists of four disjoint regions, marked as $T_i, i=1,\cdots, 4$
and evaluated through a numerical calculation, while the
unconfinement domain consists on the two remaining regions. Note
that when $\alpha_0\rightarrow 0$ we recover the classification of
the $\alpha$--line obtained in [12].

\begin{figure}[ht]  
\centering 
\epsfig{file=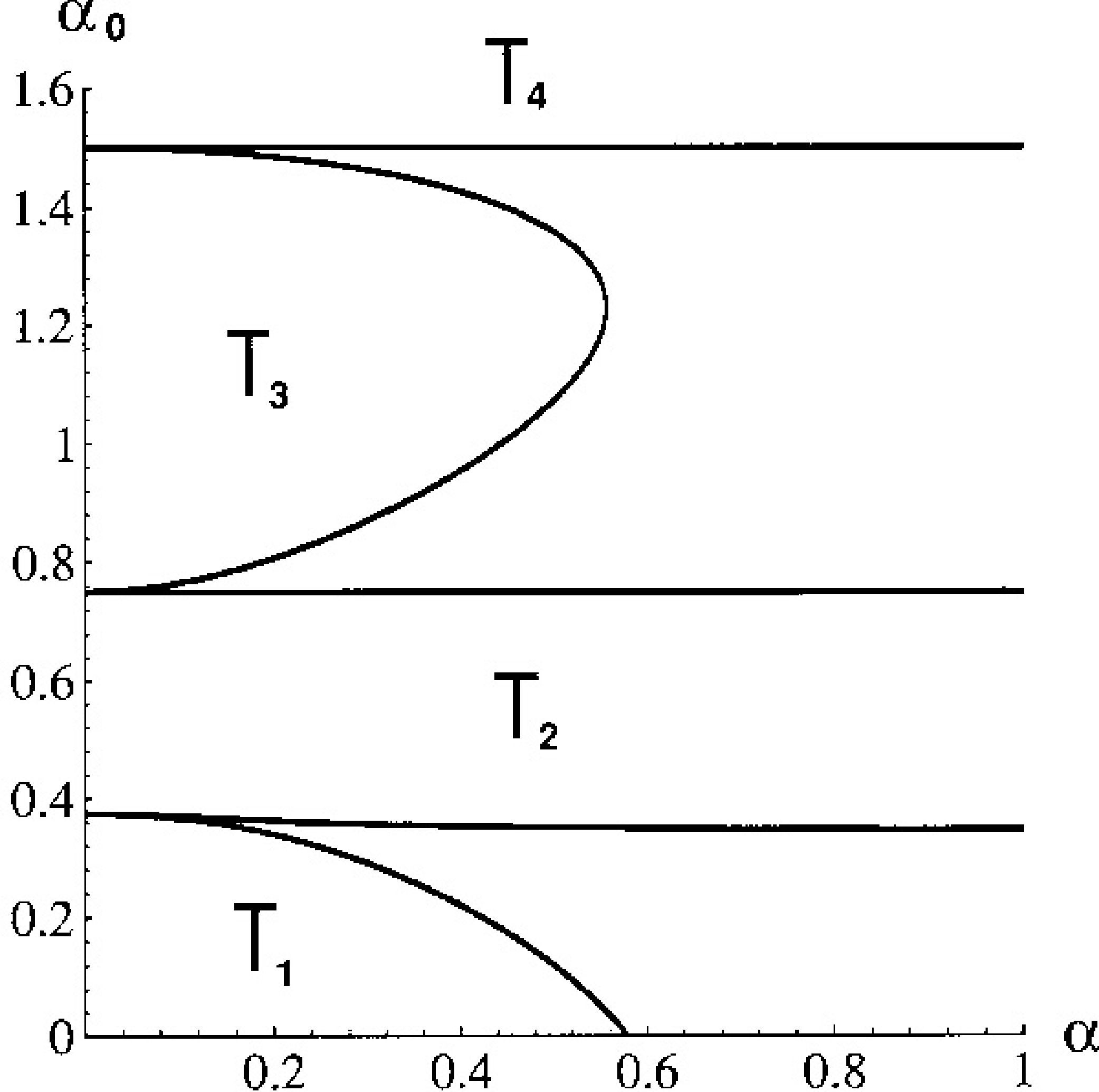, width=10cm}
\caption{\footnotesize Classification of the $\alpha-\alpha_0$ plane
according to the confined or unconfined canonical trajectories
induced by $e^{\Lambda t}$ at each point $(\alpha,\alpha_0)$ for a
charged spinless particle evolving in the fields produced by the
Penning loop with $w=4\alpha_0/3$ plus a rotating magnetic field.
The confinement domain consists of the four disjoint regions $T_i,
i=1,\cdots,4$ while the remaining regions are the
unconfinement domain.}
\end{figure}

Suppose the parameters $(\alpha,\alpha_0)$ lie in some of the
$T_j$--regions of Fig.1. Therefore, the eigenvalues of $\Lambda$
take the form $\{\pm i\omega_1,\pm i\omega_2,\pm i\omega_3\}$, and
the Hamiltonian in the rotating frame can be decomposed as
[9,12]:
$$
G=\sum_{i=1}^3\epsilon_i\omega_i\left(A_i^{\dag}A_i+
{{\epsilon_i}\over 2}\right), \eqno(5)
$$
where $A_i^{\dag}, \ A_i$ are the ladder operators satisfying 
$[A_i, A_j^{\dag}]=\epsilon_j \delta_{ij}$, and $\epsilon_i= \pm 1$. Note that 
when $\epsilon_i=1$, the $A_i^{\dag}$ and $A_i$ are equivalent to the standard 
creation and annihilation operators respectively, while when $\epsilon_i=-1$ 
the correspondence is in the inverted order. Expressions (4--5) suggest that 
the eigenstates of $G$, denoted by $|n_1,n_2,n_3\rangle$ and associated to the 
eigenvalues ${\cal E}_{n_1n_2n_3}$, form a natural base of bounded states of 
the system. Suppose that $|\psi(0)\rangle =|n_1,n_2,n_3\rangle$, then we can 
see that the state $|\psi(t)\rangle=U(t)|\psi(0)\rangle$ is cyclic with period 
$T=2\pi/\omega$: 
$$
|\psi(T)\rangle=e^{-i{\cal
E}_{n_1n_2n_3}T}|\psi(0)\rangle. \eqno(6)
$$
A direct calculation (see Eq. (1)) shows that the geometric AA phase
associated to this cyclic motion takes the form:
$$
\beta_{n_1n_2n_3}=2\pi\langle n_1,n_2,n_3 |L_3|n_1,n_2,n_3 \rangle.
\eqno(7)
$$
We can go a step further by expressing $\beta_{n_1n_2n_3}$ in terms
of ${\cal E}_{n_1n_2n_3}$ [9]. A simple calculation provides:
$$
\beta_{n_1n_2n_3} =-2\pi {\partial\over{\partial\omega}} {\cal
E}_{n_1n_2n_3} = -2\pi\sum_{i=1}^3\epsilon_i\left(n_i+{
1\over 2}\right){{\partial\omega_i}\over{\partial\omega}}, \eqno(8)
$$
where $n_i\in{\bf Z}^+, \ i=1,2,3$. It is important to remark that
relations (7--8) are valid even if the precession of the
magnetic field is fast. Relation (8) is the simplest generic form
for the geometric phase when the evolution operator possesses the
disentangled form (4). 

It is interesting at this point to discuss how the geometric phases (8) 
influence on the physical properties of our system (for other kind of systems 
see for instance [15--17]). To this end, take (2) as an unperturbed 
Hamiltonian and add to it a small periodic perturbation of frequency 
$\omega_p$. It is known from time-dependent perturbation theory that when 
$\omega_p \simeq{\cal E}_{n_1n_2n_3}-{\cal E}_{n'_1n'_2n'_3}$, where ${\cal 
E}_{n_1n_2n_3}={\cal E}_{n_1n_2n_3}(\omega)$, ${\cal E}_{n'_1n'_2n'_3}={\cal 
E}_{n'_1n'_2n'_3}(\omega)$, resonance phenomena will occur [15]. Suppose now 
that $\omega$ changes slightly by $\delta \omega$, i. e. $\omega \to \omega + 
\delta \omega$, then the unperturbed Hamiltonian (2) will be modified to 
$H(\omega + \delta \omega)$. Accordingly, in the perturbed case, the new 
resonance peak will become $\omega'_p \simeq {\cal E}_{n_1n_2n_3}(\omega + 
\delta \omega)-{\cal E}_{n'_1n'_2n'_3}(\omega + \delta \omega)$. Taking only 
the first two terms of the Taylor series of $\omega'_p$ with respect to 
$\delta \omega$ we have $\omega'_p \simeq \omega_p - {1 \over 
2\pi}(\beta_{n_1n_2n_3} - \beta_{n'_1n'_2n'_3})\delta \omega$.  As we can see, 
the resonance peak shifts by a quantity  which depends only on the geometric 
phases of the two "quasienergy" levels under consideration. In particular if  
$\omega \to 0$ the resonance peak shifting depends on the adiabatic (Berry) 
phase. 

For calculational goals, we can express 
${{\partial\omega_i}\over{\partial\omega}}$ in terms of $\omega_i$ itself. 
This last relation is more involved than (8) but it is useful when specific 
values of the parameters of the system are given and we want to evaluate the 
contribution of ${ {\partial\omega_i}\over{\partial\omega}}$ to the geometric 
phase for each degree of freedom in which the original problem has been 
decomposed. We determined ${{\partial\omega_i}\over{\partial\omega}}$ for the 
adiabatic case using this technique and taking the limit $\omega\rightarrow 
0$. We present the results in Fig.2, where in Fig. 2a we have plotted 
$\cos\theta=B_0(B_0^2+B^2)^{-1/2}$, closely related to  the solid angle in the 
${\bf B}(t)$--space, as a function of the ratio $k\equiv B/B_0>0$; the 
contributions ${{\partial\omega_i}\over{\partial\omega}}_{|\omega=0}, \ 
i=1,2,3$ as functions of $k$ are shown in Figs. 2b, 2c and 2d respectively. As 
we can see, $\cos\theta$ and ${{\partial\omega_1}\over{\partial\omega}}$ are 
well defined for all $k>0$ while ${{\partial\omega_i}\over{\partial\omega}}, 
i=2,3$ are defined just in the interval $0<k<k_{cr}\approx 0.25831$. This is 
so because, unlike $\omega_1$, outside that $k$--interval the two degrees of 
freedom associated to $\omega_2,\omega_3$ are not equivalent to the harmonic 
oscillator Hamiltonian. Therefore, those two degrees of freedom will produce 
unconfined motion on the canonical trajectories of our system. In that case 
there are not cyclic motions and so our relations to evaluate BP loose sense. 
Formulae (5--8) are valid for values of $k$ inside that interval, and because 
the three ${{\partial\omega_i}\over{\partial\omega}}$ cannot be just linear 
functions on $\cos\theta$, then the quadrupolar Penning potential affects them 
in a sensible way in contrast with a harmonic oscillator potential where they 
become proportional to $B_0(B_0^2+B^2)^{-1/2}$. In the oscillator case the 
adiabatic phase is a function just on the parameters in the ${\bf B}(t)$--
space, for this reason the dependence of BP on the spherical symmetry of the 
potential is not obvious. If, however, we choose a cylindrically symmetric 
potential (as the Penning trap potential of this work), the geometric phase 
will be a function on the binding parameters and, therefore, that dependence 
becomes apparent. So, in determining the BP for a bounded system we have to 
take into account both, the evolving parameters of the system and the symmetry 
of the binding potential which affects the BP through the eigenstates of the 
Hamiltonian. 

\begin{figure}[ht]
\centering \epsfig{file=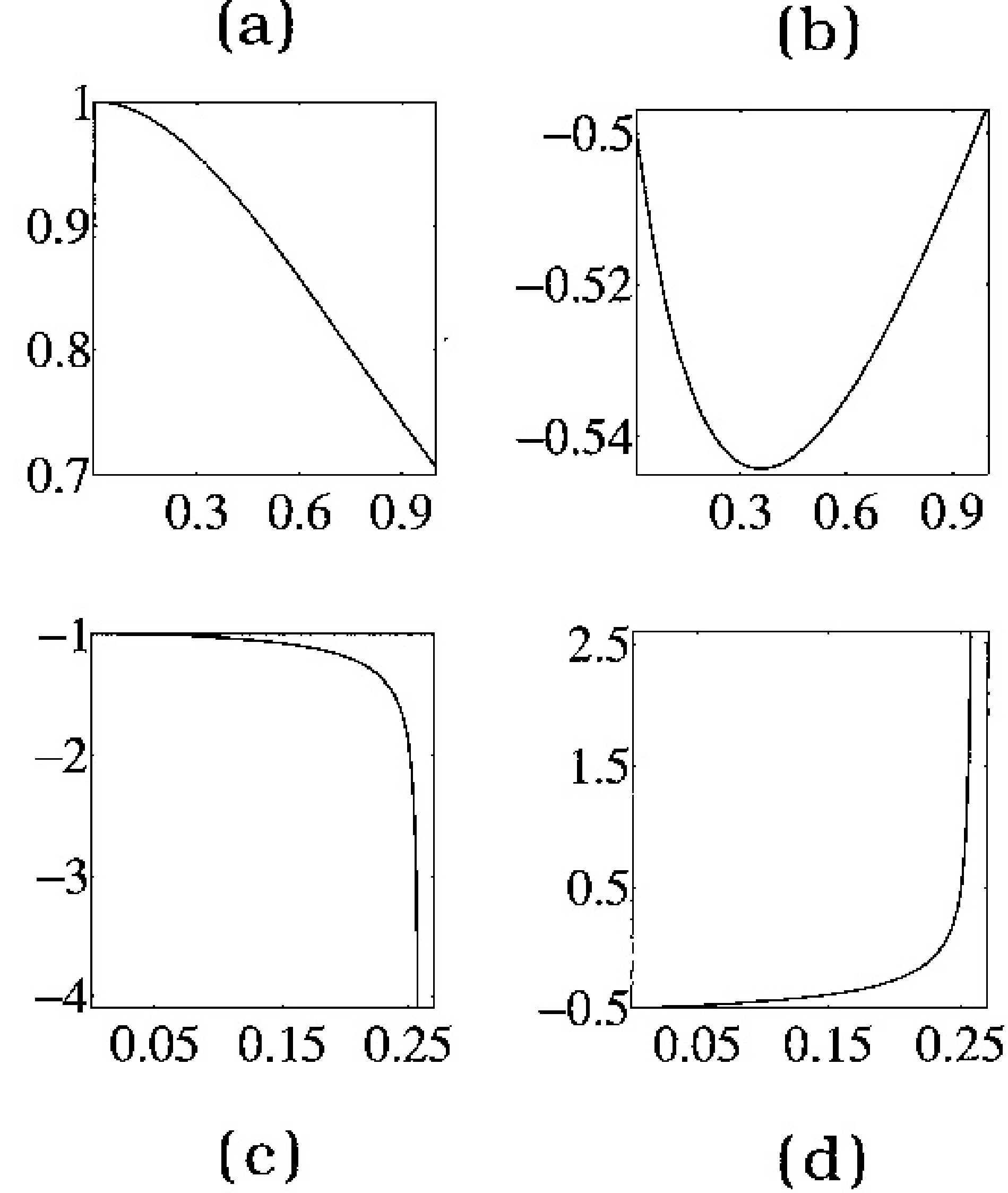, width=10cm}
\caption{\footnotesize The relevant quantities for the Berry phase 
${{\partial\omega_i}\over{\partial\omega}}_{|\omega=0}, \ i=1,2,3$ as 
functions of $k=B/B_0$ are shown for two binding quadratic potentials 
having different symmetries. For a 3--dimensional oscillator (spherical 
symmetry) all three ${ 
{\partial\omega_i}\over{\partial\omega}}_{|\omega=0}, \ i=1,2,3$ are 
proportional to $\cos\theta=(1+k^2)^{-1/2}$; in 2a we plot $\cos\theta$ 
versus $k$. For the Penning loop with $w=4\alpha_0/3$ (cylindrical 
symmetry) we display in 2b, 2c and 2d respectively the ${ 
{\partial\omega_i}\over{\partial\omega}}_{|\omega=0}, \ i=1,2,3$ as 
functions of $k$.}
\end{figure}

As an open question remains the possibility of finding a parameter's space 
${\bf X}(t)$ such that $\alpha ({\bf X}(t)) = {\bf B}(t)$ with $\alpha$ being 
an automorphism of $R^3$, ${\bf X}(t) \in R^3$. It might be that in such a 
space the BP be again a linear function of $\cos \theta$. 

To conclude, in this paper we have determined the confinement and
unconfinement regions (see Fig.1) for the motion of a
non--relativistic charged spinless particle evolving in the
superposition of the fields produced by one of the Penning loops
plus a rotating magnetic field. Working in the confinement domain
we have calculated the AA phases and taking their adiabatic limit
($\omega\rightarrow 0$) the BP were obtained. We display graphically
the differences between the relevant quantities for the BP associated
to each degree of freedom in which the motion is decoupled and the
corresponding ones obtained for a spherically symmetric binding
potential. So, in Fig.2 the contrast between the two different
symmetries is clearly shown, as might be expected for a quantity
containing some of the geometric properties of the system.

\bigskip

D.J.F.C. acknowledges financial support from the Instituto de
Cooperaci\'on Iberoame\-ricana of the Agencia Espa\~nola de
Cooperaci\'on Internacional (Spain). N.B. acknowledges financial
support from Direcci\'on General de Investigaci\'on Cient\'{\i}fica y
T\'ecnica (Spain).

\end{document}